\documentclass[11pt]{article}
\usepackage{lmodern}
\usepackage{arxiv}
\usepackage[utf8]{inputenc}
\usepackage[T1]{fontenc}
\usepackage{hyperref}
\usepackage{url}
\usepackage{booktabs}
\usepackage{amsfonts}
\usepackage{nicefrac}
\usepackage{microtype}
\usepackage{lipsum}
\usepackage{fancyhdr}
\usepackage{graphicx}
\usepackage[numbers]{natbib}
\usepackage{doi}
\usepackage{tcolorbox}

\title{The Genetic Code Paradox: Extreme Conservation Despite Demonstrated Flexibility}

\author{
  Marc Bara Iniesta \\
  \texttt{marc.bara.iniesta@gmail.com} \\
  \href{https://orcid.org/0009-0005-1480-5760}{ORCID: 0009-0005-1480-5760} \\
}

\begin{document}
\maketitle

\begin{abstract}
The universal genetic code presents a fundamental paradox in molecular biology. Recent advances in synthetic biology have demonstrated that the code is remarkably flexible—organisms can survive with 61 codons instead of 64, natural variants have reassigned codons 38+ times, and fitness costs of recoding stem primarily from secondary mutations rather than code changes themselves. Yet despite billions of years of evolution and this proven flexibility, approximately 99\% of life maintains an identical 64-codon genetic code. This extreme conservation cannot be fully explained by current evolutionary theory, which predicts far more variation given the demonstrated viability of alternatives. I propose that this paradox—evolutionary flexibility coupled with mysterious conservation—reveals unrecognized constraints on biological information systems. This paper presents testable predictions to distinguish between competing explanations: extreme network effects, hidden optimization parameters, or potentially, computational architecture constraints that transcend standard evolutionary pressures.
\end{abstract}

\keywords{genetic code \and evolution \and information theory \and synthetic biology \and biological constraints}

\section{Introduction}

The genetic code represents one of nature's most successful information processing systems, mapping 64 three-nucleotide combinations (codons) to 20 amino acids and stop signals with remarkable fidelity. From an information theory perspective, this biological system exhibits properties typically associated with optimized communication protocols: redundancy that minimizes error impact, structured organization that facilitates accurate decoding, and conservation patterns suggesting fundamental constraints on the encoding space. The code achieves approximately 2 bits of information per nucleotide position while maintaining exceptional robustness to mutations—a balance between information density and error tolerance that approaches theoretical optima for noisy channels. Yet unlike engineered systems that evolve rapidly through iterative improvement, this molecular information system has remained virtually unchanged for billions of years, with approximately 99\% of all sampled genomes maintaining identical codon assignments \citep{koonin2017origin}.

This universality might seem unremarkable—perhaps the code represents an optimal solution discovered early in evolution and maintained by natural selection. However, recent discoveries in synthetic biology have shattered this comfortable explanation. Scientists have successfully engineered viable organisms with fundamentally altered genetic codes: \textit{E. coli} strains using only 61 codons instead of 64 \citep{fredens2019}, organisms with reassigned stop codons \citep{wang2024}, and systems incorporating non-canonical amino acids \citep{chen2023}. Even more striking, when these organisms show reduced fitness, the costs stem primarily from pre-existing mutations rather than the codon changes themselves.

Nature, too, has experimented with the code. Researchers have documented over 38 natural variations across different branches of life \citep{shulgina2021}. Mitochondria use alternative codon assignments. Some fungi have reassigned the CTG codon. Certain ciliates and mycoplasmas read UGA as tryptophan instead of "stop." The code can change. It has changed. It continues to change.

This creates a profound paradox: if the genetic code is flexible enough to be rewritten in the laboratory and has been modified dozens of times in nature, why does 99\% of life stubbornly maintain the original version?

The precision of this conservation—exactly 64 codons, precisely 20 canonical amino acids—suggests constraints beyond simple biochemical requirements, potentially reflecting fundamental limits on biological information processing.

This paper proceeds as follows: Section 2 establishes the flexibility of the genetic code through synthetic biology and natural examples. Section 3 articulates the conservation paradox this flexibility creates. Section 4 examines three potential explanations, ranging from well-established (network effects) to speculative (computational constraints). Section 5 proposes specific experiments to distinguish between these explanations. Finally, we discuss the implications for evolutionary biology, synthetic biology, and our understanding of biological information processing.

\begin{tcolorbox}[title=Key Concepts]
\textbf{Codon}: A three-letter DNA sequence that codes for one amino acid\\
\textbf{64 codons}: With 4 DNA letters (A,T,G,C) in groups of 3, there are $4^3 = 64$ possible combinations\\
\textbf{Genetic code}: The mapping between these 64 codons and the 20 amino acids they encode\\
\textbf{Syn61}: An engineered E. coli that uses only 61 of the 64 possible codons
\end{tcolorbox}

\section{The Flexibility of the Genetic Code}

\subsection{Engineering New Codes: Shattering the Frozen Accident}

The traditional view of the genetic code as a "frozen accident"—unchangeable due to its deep integration into cellular machinery—has been dramatically overturned by recent synthetic biology achievements. These experiments don't merely tweak the edges of the genetic code; they fundamentally restructure it, proving that what we once thought impossible is merely difficult.

The most striking demonstration comes from the laboratory of Jason Chin at the MRC Laboratory of Molecular Biology. Fredens et al. \citep{fredens2019} accomplished what would have seemed like science fiction just two decades ago: they created Syn61, an \textit{Escherichia coli} strain with a fully synthetic genome that uses only 61 of the 64 possible codons. This monumental achievement required synthesizing the entire 4-megabase \textit{E. coli} genome from scratch, systematically recoding over 18,000 individual codons throughout the genome.

Comprehensive analysis by Nyerges et al. \citep{nyerges2024} revealed that synonymous recoding affects multiple levels of gene expression beyond simple codon replacement. The changes disrupt mRNA secondary structures, alter the positioning of regulatory motifs, and create imbalances in tRNA availability. These multi-level perturbations explain why recoded organisms require extensive adaptive evolution to regain even partial fitness, demonstrating that the genetic code's conservation stems from its deep integration into every aspect of cellular information processing.

The scope of this modification cannot be overstated. Every instance of the three eliminated codons (UAG, UAA, and AGU) had to be replaced with synonymous alternatives. Every gene had to be checked for functionality. The entire genome had to be assembled from chemically synthesized DNA fragments. Yet despite these massive changes—modifications that should have been catastrophic according to the frozen accident hypothesis—the organism lives, grows, and reproduces.

Building on this success, Wang et al. \citep{wang2024} pushed even further, creating \textit{E. coli} strains that reassigned all three stop codons for alternative functions. These "Ochre" strains don't just compress the genetic code; they repurpose it, using formerly termination signals to incorporate non-canonical amino acids. This expansion allows these organisms to produce proteins containing chemical functionalities that natural evolution has never explored—amino acids with novel reactive groups, fluorescent properties, or chemical handles for further modification.

The fitness costs of these modifications reveal perhaps the most important insight. Syn61 grows approximately 60\% slower than wild-type \textit{E. coli} under laboratory conditions—a significant but not catastrophic deficit. However, detailed genetic analysis revealed a crucial finding: the performance costs stem primarily not from the codon reassignments themselves, but from pre-existing suppressor mutations and genetic interactions that became problematic in the new genetic context \citep{wang2024}. When these secondary issues were addressed through additional engineering, fitness improved substantially.

This finding fundamentally challenges our understanding of genetic code evolution. The act of changing the code—even dramatically, affecting thousands of genes simultaneously—is not inherently deleterious. The genetic code is not frozen by intrinsic biochemical constraints but rather by the accumulation of historical contingencies that can, with sufficient effort, be overcome.

\subsection{Natural Variations: Evolution's Ongoing Experiments}

While laboratory achievements demonstrate what's possible under controlled conditions, nature provides even more compelling evidence for genetic code flexibility. Comprehensive genomic surveys, particularly the systematic screen by Shulgina and Eddy \citep{shulgina2021} analyzing over 250,000 genomes, have revealed that genetic code variations are not rare curiosities but recurring evolutionary experiments.

The documented variations span all domains of life and employ diverse molecular mechanisms:

\textbf{Mitochondrial Variations}: Perhaps the most widespread alternatives occur in mitochondria, the cellular powerhouses that maintain their own genetic systems. Vertebrate mitochondria reassign AGA and AGG from arginine to stop signals, while UGA changes from stop to tryptophan. These changes affect fundamental cellular processes—every protein made in mitochondria uses these alternative assignments. The fact that these variations exist in virtually every animal cell demonstrates that code changes can be not just tolerated but stably maintained for hundreds of millions of years.

\textbf{Nuclear Code Variations in Ciliates}: Among the most dramatic natural variations occur in ciliated protozoans. Some species reassign UAA and UAG (typically stop codons) to encode glutamine. This change requires coordinated evolution of the translation termination machinery, as these organisms must use only UGA to terminate protein synthesis. The successful radiation of ciliates—they're found in virtually every aquatic environment—proves that even fundamental changes to translation termination are evolutionarily viable.

\textbf{The CTG Clade}: A group of Candida species independently evolved a remarkable change: CTG, normally encoding leucine, instead specifies serine. This change is particularly striking because leucine and serine have very different chemical properties—one is hydrophobic, the other polar. Every CTG codon in thousands of genes changed its meaning, yet these organisms thrive. Some species in this clade even maintain ambiguous decoding, with CTG translated as both leucine and serine, suggesting an evolutionary intermediate state.

\textbf{Mycoplasma and Mitochondrial UGA Reassignment}: Multiple bacterial lineages and mitochondria independently reassigned UGA from a stop codon to tryptophan. The convergent evolution of this same change suggests it may offer selective advantages under certain conditions, perhaps related to genome reduction or metabolic optimization.

These natural experiments demonstrate several crucial principles:

First, genetic code changes can and do occur throughout evolutionary history. They're not confined to ancient evolutionary transitions but continue to arise and become fixed in modern lineages. Second, the same changes have evolved independently multiple times, suggesting that certain modifications may be particularly accessible or advantageous. Third, organisms with variant codes don't occupy marginal ecological niches—they include important pathogens, symbionts, and free-living species across diverse environments.

The pattern of natural variations also reveals constraints. Most changes affect codons that are rare in the organisms that reassign them, minimizing the number of genes that must be compatible with the new assignment. Stop codon reassignments are particularly common, perhaps because they affect fewer genes than sense codon changes. Yet even with these constraints, the existence of 38+ documented variations across the tree of life definitively refutes any claim that the genetic code cannot change.

\subsection{Mechanisms of Flexibility: How Codes Change}

Understanding how genetic codes can change provides crucial insight into why such changes remain rare despite being possible. Research has identified multiple molecular mechanisms that enable code evolution, each with distinct evolutionary dynamics and constraints \citep{ling2015}.

\textbf{Codon Capture}: The most straightforward mechanism occurs when a codon becomes rare or entirely absent from a genome. If no genes use a particular codon, it can be reassigned without fitness cost. This appears to explain many stop codon reassignments—in small genomes under pressure for compaction, one or more stop codons may disappear entirely through genetic drift. The translation machinery can then evolve new functions for these "free" codons. Modern genome engineering exploits this principle, first eliminating target codons through synonymous substitutions before reassigning them.

\textbf{Ambiguous Intermediate States}: Contrary to intuition, genetic code changes need not be binary switches. Many organisms maintain ambiguous decoding where a single codon can be translated as multiple amino acids. In some Candida species, CTG codons are decoded as both leucine and serine, with the ratio varying by growth conditions. This ambiguity creates an evolutionary bridge—organisms can explore the fitness landscape of a new code while maintaining compatibility with the old one. Such intermediates may persist for millions of years, suggesting that genetic code evolution can be gradual rather than catastrophic.

\textbf{tRNA Evolution and Modification}: The genetic code is implemented through tRNAs that physically bridge codons and amino acids. Changes to tRNA sequences, particularly in their anticodon regions, can alter codon recognition patterns. Even more subtly, post-transcriptional modifications to tRNA nucleotides can shift their specificity. Over 100 different chemical modifications have been identified in tRNAs, creating a rich landscape for evolutionary experimentation. A single nucleotide change or modification can potentially reassign multiple codons simultaneously.

\textbf{Aminoacyl-tRNA Synthetase Evolution}: These enzymes charge tRNAs with their cognate amino acids, serving as another control point for code evolution. Changes in synthetase specificity can globally alter the genetic code without any changes to tRNA sequences. Some organisms have evolved editing domains in their synthetases that allow them to discriminate between chemically similar amino acids, while others have relaxed specificity that creates natural code ambiguity.

\textbf{Release Factor Modifications}: Translation termination provides a particularly flexible target for code evolution. Unlike sense codons that require tRNA recognition, stop codons are recognized by protein release factors. These proteins can evolve more rapidly than RNAs, and changes to their recognition domains can reassign stop codons to sense codons or vice versa. The evolution of variant release factors appears to be a common route for stop codon reassignment.

Modern biotechnology has developed powerful tools that exploit these natural mechanisms at unprecedented scales:

\textbf{MAGE (Multiplex Automated Genome Engineering)}: This technique allows simultaneous modification of thousands of genetic loci, enabling systematic codon replacement across entire genomes. MAGE can generate billions of genetic variants in parallel, exploring vast evolutionary landscapes in days rather than millennia.

\textbf{CAGE (Conjugative Assembly Genome Engineering)}: By combining engineered DNA segments through bacterial conjugation, CAGE enables construction of fully synthetic genomes with arbitrary genetic codes. This approach built the Syn61 strain and continues to push the boundaries of code manipulation.

\textbf{Orthogonal Translation Systems}: Researchers have created cellular subsystems that operate with their own genetic codes independent of the host's primary code. These orthogonal ribosomes, tRNAs, and synthetases allow exploration of novel genetic codes without disrupting essential cellular functions.

The existence of these diverse mechanisms reveals that genetic code evolution faces no fundamental biochemical barriers. The translation machinery exhibits remarkable plasticity, with multiple independent components capable of evolutionary modification. The question shifts from "Can the genetic code change?" to "Given all these mechanisms for change, why does it change so rarely?"

\section{The Conservation Paradox}

Given this demonstrated flexibility, the extreme conservation of the genetic code becomes paradoxical. Consider the numbers:

- Life has existed for approximately 3.5 billion years
- Organisms undergo countless generations with mutation rates of $10^{-8}$ to $10^{-10}$ per base per generation  
- We have documented that code changes are possible and survivable
- Yet 99\% of organisms use exactly the same 64-codon system

This conservation extends beyond mere codon number. The specific assignment of codons to amino acids remains virtually identical across domains of life separated by billions of years of evolution. Why?

\subsection{Inadequacy of Current Explanations}

The traditional explanation—Crick's "frozen accident" hypothesis—proposes that once established, the code became too integrated into cellular machinery to change \citep{koonin2017frozen}. But this explanation fails to account for our observations:

\begin{enumerate}
\item \textbf{If truly frozen}: We shouldn't see any variants. Yet we see 38+.
\item \textbf{If change is catastrophic}: Engineered organisms shouldn't survive. Yet they do.
\item \textbf{If integration is the barrier}: All changes should be equally difficult. Yet some codons change more easily than others.
\end{enumerate}

The frozen accident hypothesis explains why change is difficult, not why it's so rare given that it's possible.

\subsection{The Core Questions}

This paradox raises fundamental questions:

\begin{enumerate}
\item Why exactly 64 codons? Not 61, not 67, but precisely $4^3 = 2^6$?
\item Why do organisms maintain all 64 even when they could function with fewer?
\item What force maintains this uniformity across domains of life that otherwise show enormous diversity?
\item Why have billions of years of evolution not explored more of the viable alternatives we can create in the lab?
\end{enumerate}

\section{Potential Explanations}

Several hypotheses might explain this paradox:

\subsection{Extreme Network Effects}

Perhaps the genetic code exists within a network of molecular interactions so dense that any change, while possible, incurs subtle fitness costs across thousands of processes. These individually small costs might sum to create an evolutionary valley too deep to cross under natural conditions. Laboratory organisms succeed only because we maintain them in artificial environments without competition.

Recent theoretical work has elucidated the specific mechanisms underlying these network effects. Carter and Wills \citep{carter2017} demonstrated that the genetic code exhibits reflexivity, where aminoacyl-tRNA synthetases both determine and are determined by codon assignments, creating a self-referential network that resists change. This interdependence means that any alteration propagates through multiple system levels, disrupting not only translation but also regulatory networks built upon the code. Furthermore, Koonin and Novozhilov \citep{koonin2017origin} showed how horizontal gene transfer in early microbial communities intensified selection for a common code, as only organisms with compatible codes could effectively exchange genetic material, creating a powerful convergence mechanism.

\subsection{Hidden Optimization}

The standard genetic code might optimize for parameters we haven't yet recognized. Beyond error minimization and chemical similarity, it might maximize:
\begin{itemize}
\item Information processing speed under cellular constraints
\item Robustness to specific types of environmental stress
\item Compatibility with unknown cellular processes
\item Some combination of factors that creates a unique global optimum
\end{itemize}

Recent studies have revealed multiple optimization parameters beyond error minimization that contribute to the SGC's dominance. Shenhav and Zeevi \citep{shenhav2020} discovered that the genetic code minimizes resource costs, reducing the probability that mutations produce proteins with unnecessarily high nitrogen or carbon content—a metabolic optimization independent of error minimization. Wnętrzak et al. \citep{wnetrzak2018} employed eight-objective evolutionary algorithms to demonstrate that the SGC represents a complex compromise among multiple physicochemical properties including hydrophobicity, molecular volume, polarity, and refractivity. Notably, Błażej et al.\citep{blazej2018} found that while many alternative codes could achieve lower error costs on isolated parameters, they do so at the expense of disrupting other vital properties, with the SGC consistently occupying a deep local optimum when multiple objectives are considered simultaneously.

\subsection{Computational Architecture Constraints}

If biological systems operate within computational constraints imposed by the nature of reality itself, the genetic code might reflect fundamental limits on information processing. The number 64 equals $2^6$, a power of two ubiquitous in digital systems. This is significant because information systems often naturally organize around powers of two for efficiency—just as computer memory comes in sizes like 64, 128, or 256 megabytes, not 65 or 130. If biological information processing faces similar constraints, we might expect the genetic code to "snap" to these mathematically convenient values rather than arbitrary numbers. This might not be coincidence but necessity—a reflection of underlying computational architecture that life must respect.

Beyond evolutionary and biochemical constraints, fundamental limits on biological information processing may restrict genetic code variation. Taipale \citep{taipale2018} quantified these limits, showing that finite genomic capacity constrains the range of viable genetic codes. The genome functions as a limited communication channel where the total information content restricts how much detail can be specified about molecular interactions, forcing reliance on coarse-grained regulatory circuits. This perspective aligns with information-theoretic principles: the genome cannot specify every molecular interaction detail, and mutations represent noise in the transmitted message. Changes that would excessively perturb this delicate information balance are purged by selection, effectively limiting the evolutionary accessibility of alternative codes.

This hypothesis makes specific, testable predictions that distinguish it from purely evolutionary explanations.

\section{Testable Predictions}

Before examining specific tests, we note that these three explanations rest on different levels of empirical support. Network effects and hidden optimization parameters are well-documented through quantitative studies \citep{carter2017,shenhav2020,wnetrzak2018}, while the computational architecture hypothesis, though consistent with observed patterns, remains more speculative. The following experiments are designed to distinguish between all three possibilities, potentially validating or refuting the computational constraints hypothesis while deepening our understanding of the established mechanisms.

\subsection{Testing Network Effects}

\textbf{Prediction}: If network effects maintain code universality, then:
\begin{itemize}
\item Fitness costs should scale with organism complexity
\item Simple organisms should tolerate changes better than complex ones
\item Costs should be distributed across many cellular processes
\end{itemize}

\textbf{Test}: Compare code modifications across organisms of varying complexity, from minimal bacteria to complex eukaryotes. Measure fitness impacts in competitive environments versus laboratory conditions.

The reflexivity identified by Carter and Wills \citep{carter2017} makes a specific prediction: fitness costs should scale non-linearly with the number of altered aaRS-tRNA pairs, as each change compounds through the self-referential network.

\subsection{Testing Hidden Optimization}

\textbf{Prediction}: If the code optimizes for unrecognized parameters, then:
\begin{itemize}
\item The standard code should excel at tasks we haven't tested
\item Variant codes should show specific weaknesses
\item These advantages should be measurable once identified
\end{itemize}

\textbf{Test}: Systematically compare standard versus variant codes across diverse challenges: temperature extremes, radiation exposure, rapid environmental changes, information processing demands.

Building on the multi-objective optimization framework of Wnętrzak et al. \citep{wnetrzak2018} and the resource conservation principle of Shenhav and Zeevi \citep{shenhav2020}, we can test whether the standard code excels at previously unmeasured parameters such as translation speed under resource limitation or protein stability under temperature stress.

\subsection{Testing Computational Constraints}

\textbf{Prediction}: If computational architecture constrains the code, then:
\begin{itemize}
\item Performance should show discontinuous drops at specific mathematical boundaries
\item Organisms with 63 or 65 codons should fail differently than those with 60 or 68
\item Error patterns should reflect information-theoretic rather than chemical constraints
\end{itemize}

\textbf{Test}: Engineer organisms with codon numbers at and between powers of 2. Look for:
\begin{itemize}
\item Performance cliffs at $2^n$ boundaries (32, 64, 128)
\item Error patterns matching binary corruption rather than chemical misfolding  
\item Processing speeds that hit hard limits regardless of selection pressure
\end{itemize}

Santos and Monteagudo \citep{santos2017} demonstrated that the fitness landscape of genetic codes is highly rugged, with the SGC residing in a deep local optimum. If computational constraints shape this landscape, we should observe that codes deviating from powers of 2 in their state space (e.g., 63 or 65 codons) encounter disproportionate fitness penalties compared to those maintaining binary-compatible architectures.

\subsection{The Translation Speed Limit Test}

Current data shows translation speeds of:
\begin{itemize}
\item Bacteria: 10-21 amino acids/second
\item Eukaryotes: 3-10 amino acids/second
\item Thermophiles: up to 25 amino acids/second
\end{itemize}

\textbf{Key test}: Can any organism, under any conditions, exceed approximately 35 amino acids/second? If this represents a hard computational limit rather than a chemical constraint, no amount of evolutionary pressure should break it.

\section{Implications}

Resolving this paradox will have concrete implications for both fundamental biology and synthetic biology applications.

For evolutionary biology, understanding why the genetic code remains frozen despite demonstrated flexibility will reveal the true nature of evolutionary constraints. If network effects are responsible, it suggests that molecular evolution operates within far narrower viable spaces than current models predict. This would require revising our understanding of evolutionary accessibility and the relationship between genotype and phenotype spaces.

For synthetic biology, identifying the source of genetic code conservation is crucial for engineering organisms with radically altered codes. If the constraint is network-based, successful engineering will require systematic rewiring of cellular networks beyond just codon reassignment. If hidden optimization parameters exist, discovering them could guide the design of novel genetic codes optimized for specific applications: enhanced incorporation of non-canonical amino acids, improved resistance to viral infection, or specialized protein production systems.

For information theory and the origin of life, this paradox touches on fundamental questions about biological information processing. The specific quantization at 64 states ($4^3$) and 20 amino acids may reflect optimal solutions to the dual challenges of information storage and error correction under prebiotic constraints. Understanding these principles could inform theories about how the first genetic systems emerged and why they converged on this particular architecture.

Finally, the methodological approach proposed here—using synthetic biology as a tool to probe evolutionary constraints—establishes a new paradigm for investigating biological universals. By engineering systems that violate apparent rules and carefully analyzing their failure modes, we can distinguish between true constraints and historical accidents. This approach could be applied to other biological universals: Why do all cells use ATP? Why is the central dogma unidirectional? What other biological features appear immutable but might simply reflect unexplored evolutionary space?

\section{Conclusion}

The genetic code presents us with a genuine scientific paradox. We know the code can change: we've changed it ourselves and nature has changed it dozens of times. We know these changes can be tolerated: organisms survive and reproduce with altered codes. Yet 99\% of life maintains exactly the same system, suggesting a constraint we don't understand.

This paradox cannot be dismissed with hand-waving about frozen accidents or metabolic integration. It demands explanation. The tests proposed here can distinguish between evolutionary, informational, and computational explanations for this mysterious conservation.

The resolution of this paradox may require perspectives beyond traditional molecular biology. In information theory and signal processing, we recognize that certain system architectures emerge not from local optimization but from fundamental constraints on information transmission and processing. The genetic code's strict adherence to 64 states ($4^3$) despite demonstrated flexibility at 61 or other values mirrors how certain quantization levels emerge naturally in information systems, not by design preference but due to underlying architectural efficiencies. Similarly, the translation machinery's apparent speed limits across all domains of life suggest possible information-processing bottlenecks rather than purely biochemical constraints.

Based on current evidence, network effects combined with multi-parameter optimization provide the most empirically supported explanation for genetic code conservation. The reflexivity of the translation system \citep{carter2017} and the multi-objective optimization revealed by evolutionary algorithms \citep{wnetrzak2018} demonstrate concrete mechanisms maintaining the code. However, these mechanisms do not fully explain the precise convergence on 64 codons rather than neighboring values. Whether this precision reflects additional computational constraints or simply the mathematical inevitability of a triplet code with four bases remains an open question—one that the experiments proposed here could help resolve.

Whatever the answer, resolving this paradox will reveal fundamental principles about how information systems evolve and the constraints they face, whether those constraints emerge from biochemistry, information theory, or something deeper still.

\bibliographystyle{unsrt}

\end{document}